\documentclass[%
aps,
prb,
superscriptaddress,
 sd,%
 amsmath,amssymb,
 reprint,%
]{revtex4-1}

\usepackage{graphicx}% Include figure files
\usepackage{amsmath}
\usepackage[]{natbib,hyperref}
\hypersetup{colorlinks=true,allcolors=blue}
\usepackage{todonotes}
\usepackage[version=3]{mhchem} % Formula subscripts using \ce{}
\usepackage[normalem]{ulem}

\begin{document}

\title{Layer- and Frequency-Dependent Second Harmonic Generation in Reflection from GaSe Atomic Crystals}
\author{Yanhao Tang}
\affiliation{Department of Physics and Astronomy, Michigan State University, East Lansing, MI 48824, USA}
\author{Krishna C. Mandal}
\affiliation{Department of Electrical Engineering, University of South Carolina, Columbia, SC 29208, USA}
\author{John A. McGuire}
\email{mcguire@pa.msu.edu}
\affiliation{Department of Physics and Astronomy, Michigan State University, East Lansing, MI 48824, USA}
\author{Chih Wei Lai}
\email{cwlai@msu.edu}
\affiliation{Department of Physics and Astronomy, Michigan State University, East Lansing, MI 48824, USA}
\affiliation{Materials Sciences Division, Lawrence Berkeley National Laboratory, Berkeley, CA 94720, USA}
\affiliation{US Army Research Laboratory, Adelphi, MD 20783, USA}
%\date{\today}

\begin{abstract}
We report optical second-harmonic generation (SHG) in reflection from GaSe crystals of 1 to more than 100 layers using a fundamental picosecond pulsed pump at 1.58 eV and a supercontinuum white light pulsed laser with energies ranging from 0.85 to 1.4 eV. The measured reflected SHG signal is maximal in samples of $\sim$20 layers, decreasing in thicker samples as a result of interference. The thickness- and frequency-dependence of the SHG response of samples thicker than $\sim$7 layers can be reproduced by a second-order optical susceptibility that is the same as in bulk samples. For samples $\lesssim$7 layers, the second-order optical susceptibility is reduced compared to that in thicker samples, which is attributed to the expected bandgap increase in mono- and few-layer GaSe.
\end{abstract}

\maketitle
\section{Introduction}

Two-dimensional semiconductors such as graphene and transition metal dichalcogenides (TMDs, e.g., \ce{MoS2}, \ce{WS2}, and \ce{WSe2}) have been studied intensively as potential materials to complement silicon electronics and gallium arsenide optoelectronics. A recent trend in the search for new two-dimensional systems is the isolation and study of atomically thin sheets of layered materials. As in the case of graphene and TMDs, gallium monochalcogenide nanoslabs from atomic thickness to hundreds of layers can be prepared by mechanical exfoliation. In contrast to TMDs, gallium monochalcogenides (MXs) either have a direct bandgap or nearly degenerate indirect and direct bandgaps in bulk, making them versatile materials in which strong emission occurs and light-matter coupling and spin polarization can be controlled from bulk to atomically thin crystals \cite{tang2015,tang2015a,tang2015b}. For example, the unique band structure of GaSe allows generation and preservation of a high degree of spin polarization for both electrons and holes \cite{gamarts1977,ivchenko1977,tang2015,tang2015a,tang2015b,do2015,li2015}.

\begin{figure}[htbp]
	\includegraphics[width=0.9\columnwidth]{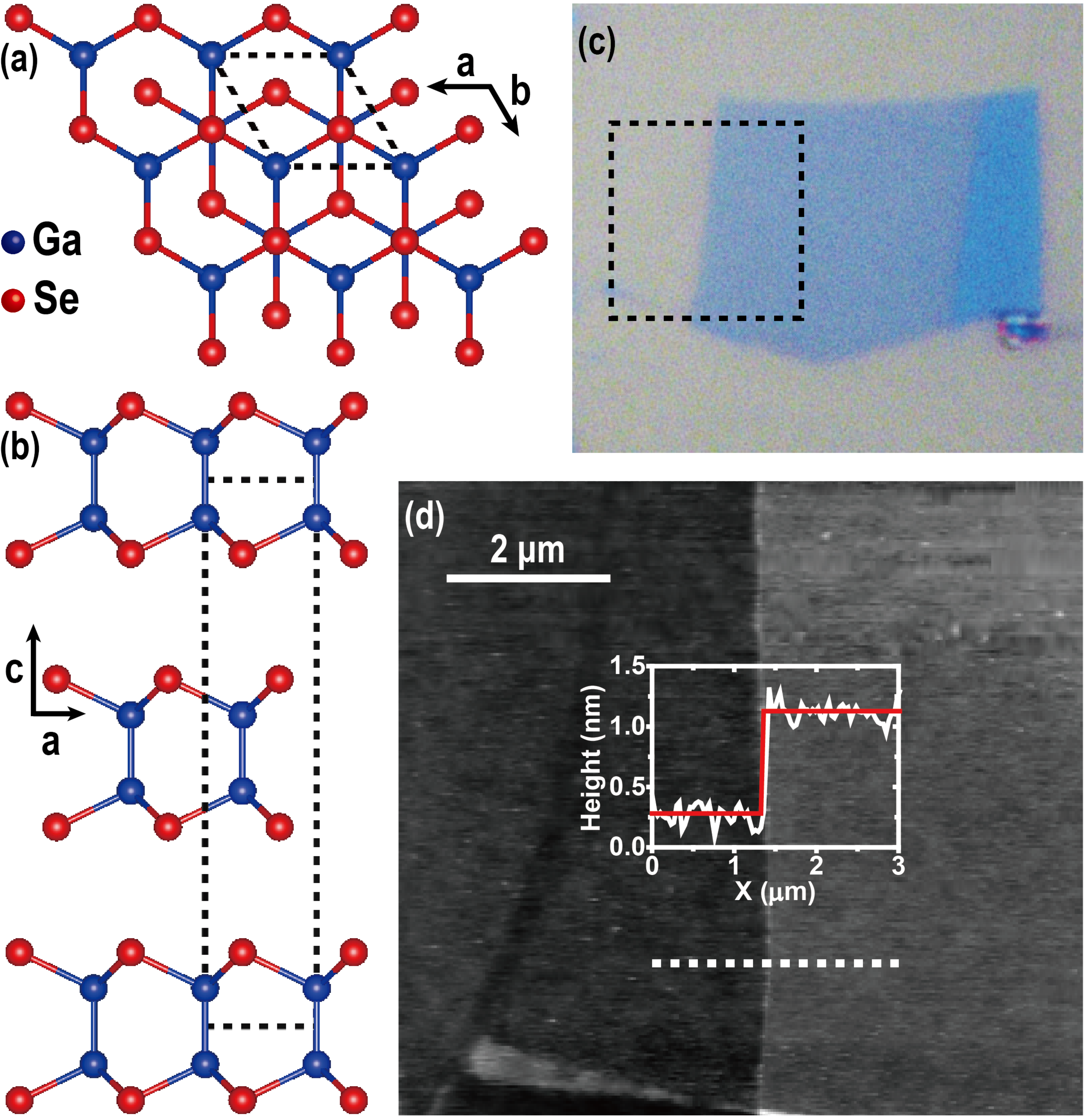}
	\caption{Optical and AFM images of monolayer GaSe. (a) Top view and (b) side view of schematic crystal structure of $\epsilon$-GaSe (ABA pattern) (c) Optical image of monolayer GaSe deposited onto Si substrate with a 90 nm SiO$_{2}$ layer. Dashed square represents the AFM region, shown in (d). (d) AFM image reveals 0.85$\pm$0.1~nm thickness of monolayer GaSe, along the dashed line.
		}\label{fig1}
\end{figure}

GaSe crystals are composed of covalently bonded layers each of which contains two gallium sublayers sandwiched by two selenium sublayers (Fig. \ref{fig1}a--b). The layers are stacked along the $c$-axis of the crystal and form several polytypes with different stacking orders. $\epsilon$-GaSe belongs to the non-centrosymmetric space group $D_{3h}$ and has been a source of long-standing interest due to its high nonlinear optical susceptibility. The second-order optical susceptibility is comparable to or greater than more commonly used nonlinear crystals such as $\beta$-BBO \cite{fernelius1994}. However, practical implementation of GaSe remains limited to the generation of THz and infrared radiation because of its softness (Mohs hardness $\approx$ 0 \cite{dmitriev2013}) and optical absorption above the bandgap. GaSe is a semiconductor with a quasi-direct bandgap of $\sim$2.0 eV. The valence band maximum is at the $\Gamma$ point, and the conduction band minimum is near the $M$ point but only about 10~meV below the local minimum at the $\Gamma$ point \cite{aulich1969, mooser1973, schluter1976, sasaki1981, capozzi1993}. Atomically thin GaSe crystals are expected to exhibit an increasing bandgap ($\sim$3.8 eV in monolayer GaSe \cite{zhuang2013}) and undergo a direct-to-indirect-bandgap transition when the top of the uppermost valence band moves away from the $\Gamma$ point in single and few-layer crystals \cite{zhuang2013,do2015,li2015}. Absorption measurements of nanoscale GaSe particles were interpreted in terms of such a size-dependent shift in the bandgap \cite{rybkovskiy2011}. 

SHG is a powerful tool for probing symmetries and electronic structure \cite{wang2015}. The efficiency of SHG in \ce{WSe2} at low temperature was shown to be enhanced by about three orders of magnitude when the SHG photon energy was in resonance with the 1s exciton peak \cite{wang2015}. The search for a layered material with quasi-direct bandgap and with broken inversion symmetry at arbitrary thicknesses motivates an exploration of the second-order optical response of GaSe from the monolayer to the bulk. Unlike the case of TMDs, in which SHG is only efficient for odd number of layers \cite{kumar2013,li2013,malard2013}, $\epsilon$-GaSe crystals remain noncentrosymmetric ($D_{3h}$) independent of number of layers, allowing for SHG in an arbitrary number of layers. Here we report on measurements of the second-harmonic response of exfoliated GaSe using a large range of sub-bandgap fundamental photon energies with corresponding second-harmonic photon energies spanning from 0.3 eV below to 1.0 eV above the bulk bandgap. By accounting for the different contributions to the second-harmonic response and wavelength-dependent interference, we are able to reproduce the frequency- and thickness-dependence of the SHG signal from samples from $\sim$7 to $\sim$100~L with a susceptibility $\big|\chi^{(2)}\big|= 80\pm18$~pm/V, similar to the reported bulk value of $\big|\chi^{(2)}\big|=2\big|d_{22}\big|=108\pm21.6$~pm/V \cite{dmitriev2013}. However, for crystals $\lesssim$7~L, we observe a suppression of the nonlinear susceptibility by as much as a factor of $\sim$5 at 3 layers. The latter observation is qualitatively consistent with a report on SHG from mechanically exfoliated GaSe crystals from 2 to 10 layers thick excited with a fundamental photon energy of 1.55~eV \cite{jie2015}. Contrary to a recent report of an enhanced $\chi^{(2)}$ in monolayer CVD-grown GaSe \cite{zhou2015}, we do not observe an increase in the efficiency of SHG or $\chi^{(2)}$ for monolayer GaSe.

\section{\label{sec:method}Methods}
Atomically thin GaSe crystals are mechanically exfoliated from a Bridgman-grown $\epsilon$-GaSe crystal \cite{mandal2008a} and deposited onto a Si substrate with a 90 nm \ce{SiO2} layer. We identify single- and few-layer GaSe crystals using an optical microscope and determine their thickness using an atomic force microscope (AFM) (Fig.~\ref{fig1}). In Fig.~\ref{fig1}c, we show the optical image of a 1~L nanoslab with size about 10 $\mu$m. The corresponding AFM measurement (Fig.~\ref{fig1}d) gives a thickness of $0.85\pm0.1$ nm, which is consistent with previous studies\cite{zhou2015} of CVD-grown monolayer GaSe. The thickness of 2~L and 3~L nanoslabs (not shown) is about an integer multiple of that of 1~L, around 1.5$\pm0.1$ nm and 2.4$\pm0.1$ nm, respectively.

SHG in reflection is generated with a fundamental pump wavelength of $\lambda=785$~nm (1.58 eV) from a 2 ps pulsed Ti:Sapphire oscillator (Coherent Mira 900D) or $\lambda>800$ nm ($<1.55$~eV) from a supercontinuum white-light laser (NKT Photonics SuperK EXTREME EXB-6). A reflective microscope objective with numerical aperture NA = 0.5 is used to focus the fundamental pump laser beam and to collect the SHG signal in the reflection geometry. The samples are maintained in vacuum ($10^{-5}$~Torr) to minimize degradation from oxidation or water contamination.

\begin{figure}[htbp]
	\includegraphics[width=0.9\columnwidth]{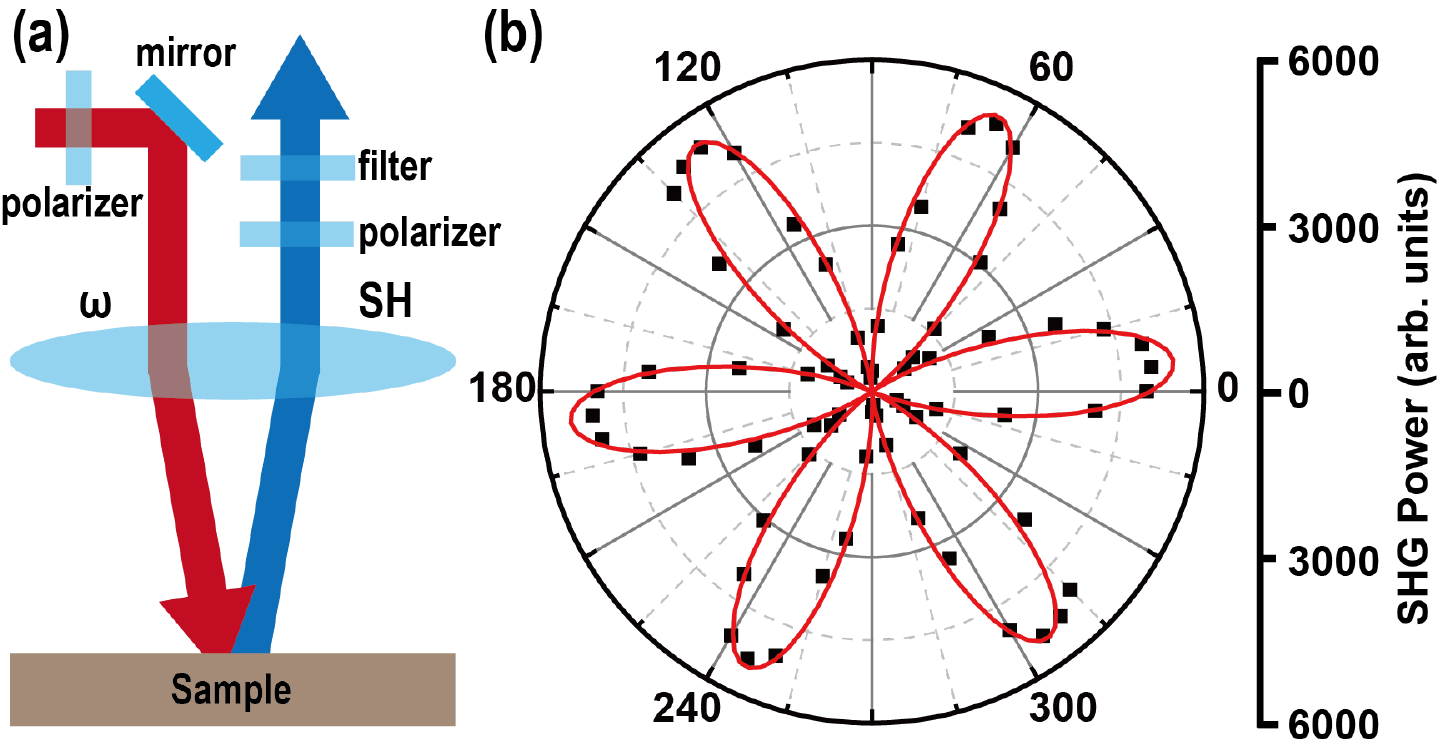}
	\caption{(a) Schematic of the experimental setup. (b) Polarization-dependent SHG power (solid squares) from monolayer GaSe detected with polarization perpendicular to that of the fundamental field. The angle represents the polarization orientation of the fundamental field. The data from $180^\circ$ to $355^\circ$ are the data from $0^\circ$ to $175^\circ$ shifted by $180^{\circ}$. The red curve is a fit to the form expected for the $D_{3h}$ symmetry.} \label{fig2_shg_pol}
\end{figure}

To confirm the crystalline symmetry, we measure polarized SHG by rotating the polarization of the fundamental pump with $\lambda=785$~nm in the $x$-$y$ plane and collecting the orthogonally polarized SHG signal (Fig. \ref{fig2_shg_pol}). The polarization-dependent detection efficiency of the spectroscopic and optical collection system is calibrated with a 400 nm laser. In this set of polarized SHG measurements, the fundamental pump laser beam with an incident angle of less than 8$^\circ$ and a flux of $7.5\times10^{-2}$~nJ per pulse is focused to a spot of $2.6~\mu$m radius. For layer-dependent SHG measurements, the SHG signals are averaged over four polarizations of the fundamental pump, namely $\psi$ = 0$^\circ$, 45$^\circ$, 90$^\circ$ and 135$^\circ$, where $\psi$ is the polarization angle of the fundamental with respect to the $x$-axis. We further determine the frequency-dependent SHG using the supercontinuum laser as the fundamental pump. The value of the nonlinear susceptibility of each GaSe sample is determined by comparing with the reflected SHG signal from the surface of a thick ($>$1~mm) BBO crystal under identical pump flux and spot size.

\section{\label{sec:result}Results and Discussion}
In Fig.~\ref{fig2_shg_pol}, we show the experimental polarization-dependence of the SHG for 1~L, where the SHG power is plotted as a function of $\psi$. The SHG response from the GaSe nanoslabs exhibits the six-fold angular-dependence expected for the $D_{3h}$ symmetry of $\epsilon$-GaSe. The data are fitted with the function $I_{SH}=I^{\omega}
\sin^2(3\psi+\psi_0)$, where $I_{SH}$ is the reflected SHG power. We label quantities associated with the fundamental fields with a superscript $\omega$; otherwise, the quantities are taken to be associated with the SH fields. The six-fold pattern is confirmed from few-layer to bulk. To further confirm the second-order nature of the SHG radiation, power-dependent SHG was measured on a 4~L nanoslab, yielding a power-dependence (not shown) of 
$\mathcal{P}_{SH}\propto \left(\mathcal{P}^{\omega}\right)^{1.96\pm0.01}$, where $\mathcal{P}_{SH}$ and $\mathcal{P}^{\omega}$ are the power of the SHG radiation and the fundamental field, respectively.

\begin{figure}[hbtp]
	\includegraphics[width=0.9\columnwidth]{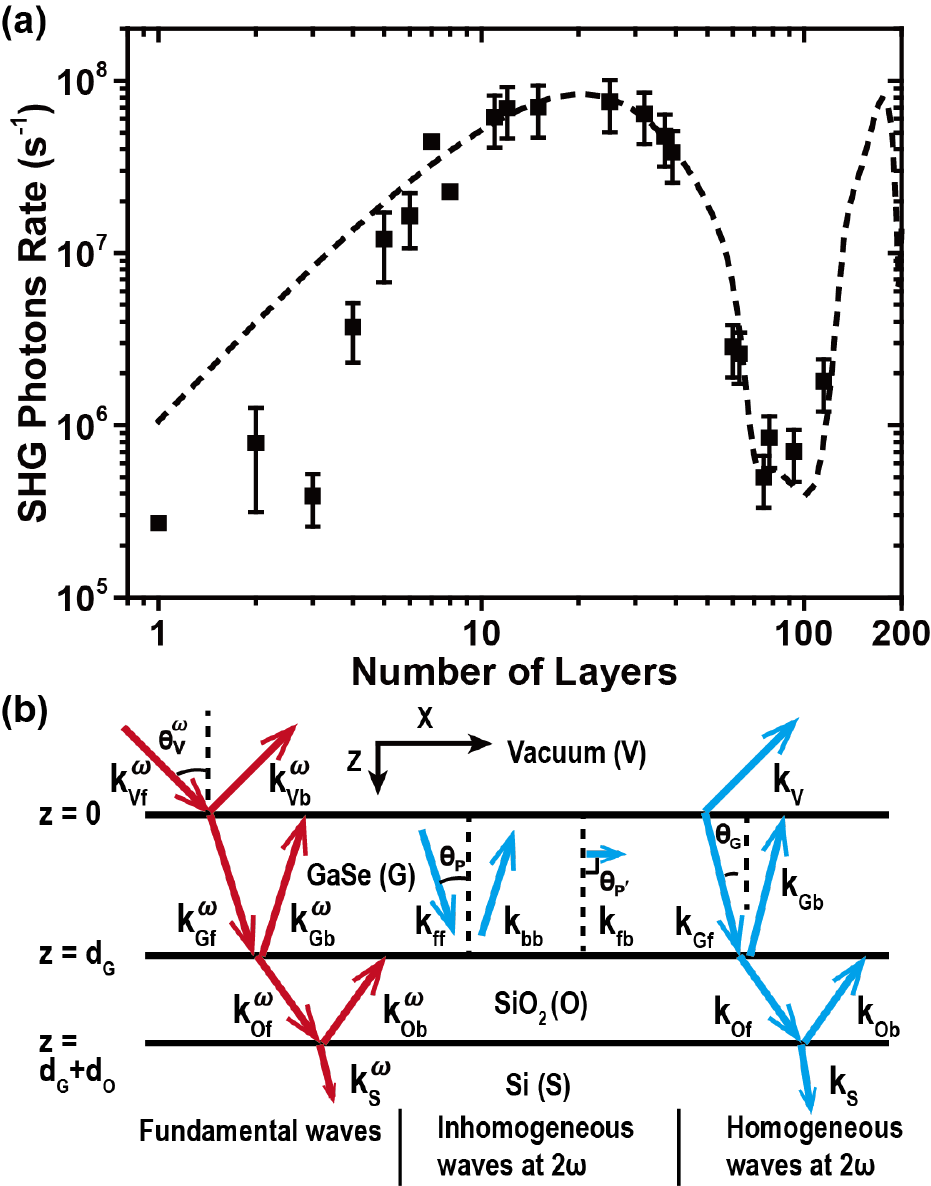}
	\caption{Layer-dependent second harmonic generation (SHG) from GaSe nanoslabs. (a) Measured rates (squares) of generation of SHG photons by GaSe nanoslabs ranging from 1~L to $>100$~L excited by a fundamental field at 785 nm. For thickness $\leq10$~L, the error bars represent the sample-to-sample variance, about $\pm 50\%$, of the average SHG power. For samples of thickness $>10$~L, which generally have a lateral size $>20~\mu$m, the error bars represent the typical position-to-position variance, about $\pm 30\%$, of the SHG power from the same nanoslab. The dashed curve is a least-squares fit to the data for thickness $\geq$ 10~L according to the model in the text, revealing a $\big|\chi^{(2)}\big|$ of $78\pm17$~pm/V. (b) A model of SHG accounting for interference in the multilayer system composed of vacuum (V), the GaSe (G) nanoslab, a 90 nm SiO$_2$ (O) layer, and a Si (S) substrate. The red and blue arrows represent the wave vectors of the fields at the fundamental and second-harmonic frequencies, respectively.
}\label{fig3_shg_layer} 
\end{figure} 

In Fig.~\ref{fig3_shg_layer}a, we show the layer-dependent SHG under a fundamental pump with $\lambda=785$~nm. In samples with thickness $d_{\textrm{G}}\geq$ 7~L, the SHG power reaches a maximum near 20~L and decreases rapidly for $d_{\textrm{G}}\geq$ 25~L. The layer-dependent SHG is reproduced with a model (Fig.~\ref{fig3_shg_layer}b) accounting for interference in the multilayer system \cite{bloembergen1962,chang1966} composed of vacuum (V), a $d_{\textrm{G}}$-thick GaSe slab (G), a $d_{\textrm{O}}$(90 nm)-thick \ce{SiO2} (O) layer, and a Si (S) substrate.

The nonlinear optical response of the GaSe slab can be understood in terms of the second-order polarization $\vec{P}^{(2)}$,
\begin{equation} 
\label{eqn:P_nls}
\vec{P}^{(2)}=\vec{P}_{ff}+\vec{P}_{bb}+2\vec{P}_{fb}
\end{equation}
where, for example, $\vec{P}_{fb}\equiv\tensor{\chi}^{(2)}:\vec{E}_{\textrm{G},f}^{\omega}\vec{E}_{\textrm{G},b}^{\omega}$, $\tensor{\chi}^{(2)}$ is the second-order nonlinear optical susceptibility, and $\vec{E}_{\textrm{G},f}^{\omega}$ and $\vec{E}_{\textrm{G},b}^{\omega}$ are the homogeneous waves at frequency $\omega$ propagating respectively in GaSe in the $+z$ (forward) and $-z$ (backward) directions. Since the depth of field (about 10 $\mu$m) is much larger than the coherence length of the SHG radiation (about 0.1 $\mu$m), we use the plane-wave approximation. For symmetry class $D_{3h}^1-\bar{6}m2$ and fundamental light along the crystalline $c$-axis, the sum of the SHG power measured with SH polarization alternately parallel and perpendicular to the fundamental is independent of sample orientation. Since the angle of incidence in vacuum is small ($\theta_{\textrm{V}}^{\omega}\equiv\theta_{\textrm{V}}\approx8^{\circ}$),  it is sufficient to calculate $\vec{P}^{(2)}$ for fundamental electric field polarization along the GaSe $x$-axis ($a$-axis), for which $\vec{P}^{(2)}=P^{(2)}\hat{y}$. 

Fitting the data for the SHG power as a function of sample thickness and wavelength amounts to solving a pair of boundary value problems. In particular, since the generation of the second-harmonic waves negligibly depletes the fundamental field, the determination of the fundamental electric field in the GaSe layer is just a standard boundary value problem for the homogeneous wave equation. For a given value of $\tensor{\chi}^{(2)}$, the fundamental field in the GaSe layer then directly yields the second-order polarization $P^{(2)}$ (i.e., the inhomogeneous waves in Fig. \ref{fig3_shg_layer}b) given by Eq. \ref{eqn:P_nls}. The electromagnetic field at $2\omega$ in each medium (i.e., the homogeneous SH waves of Fig. \ref{fig3_shg_layer}b) is then determined by solving the boundary value problem for the inhomogeneous wave equation for the electromagnetic field at $2\omega$. The SH electric and magnetic fields $E_y$ and $H_x$ (we hereafter drop the directional subscripts) at the upper and lower boundaries of medium $n=\textrm{V,G,O,S}$ are given by identical equations as for the fundamental fields except for the addition of the inhomogeneous terms in GaSe:
\begin{widetext}
\begin{eqnarray}
E_{n}&=&E_{n,f}\exp\left(i\phi_n\right)+E_{n,b}\exp\left(-i\phi_n\right)+4\pi\left[\frac{P_{ff}\exp{(i\phi_P)}+P_{bb}\exp{(-i\phi_P)}}{\epsilon_{P}-\epsilon_{n}}+\frac{2P_{fb}}{\epsilon_{P'}-\epsilon_{n}}\right]\delta_{n,G}\nonumber\\
H_{n}&=&\epsilon_n^{1/2}\cos{\theta_n}\left[E_{n,f}\exp{(i\phi_n)}-E_{n,b}\exp{(-i\phi_n)}\right]+4\pi\epsilon_{P}^{1/2}\cos{\theta_{P}} \frac{P_{ff}\exp{(i\phi_{P})}-P_{bb}\exp{(-i\phi_{P})}}{\epsilon_{P}-\epsilon_{n}}\delta_{n,G},
\end{eqnarray}
\end{widetext}
where $P_{n,ff}=P_{n,bb}=P_{n,fb}=0$ except for $n=\textrm{G}$. We define $\phi_n=\phi_{n,ff}=\phi_{n,bb} \equiv0$ at the upper surface of each medium $n=\textrm{G,O,S}$ as well as the lower surface of the vacuum. At the lower surface of each medium, $\phi_n=
\left|k_{n,z}\right|d_n=2\epsilon_n^{1/2}d_n\omega c^{-1}\cos\theta_n$~($n=\textrm{P,G,O}$), where $d_n$ is the thickness of the corresponding medium ($d_P$=$d_{\textrm{G}}$), $\theta_n$ is the angle of $\vec{k}_n$ with respect to the surface normal, and $c$ is the speed of light. In particular, the variation of $\phi_{\textrm{G}}$ with the number of GaSe layers will result in pronounced interference patterns. For the SH field, we also have $E_{\textrm{V},f}=E_{\textrm{S},b}=0$, i.e., there is no incident SH field in vacuum or reflected SH field from the far side of the thick substrate. 

The propagation directions of the homogeneous waves are determined by 
	\begin{equation}
\sin\theta_{\textrm{V}}=\epsilon_n^{1/2}\sin\theta_n~~(n=\textrm{V,G,O,S})
	\label{eqn:Snell-law} 
	\end{equation}
where $\theta_{n}$ and $\epsilon_{n}$ are respectively the refracted or reflected angles and the dielectric constant of the waves at $\omega$ or 2$\omega$ in medium $n$. For the inhomogeneous waves at 2$\omega$, the angle $\theta_P$ of $\vec{P}_{ff(bb)}$ is the same as the refracted angle of the fundamental field at $\omega$ in the GaSe slab, while the refracted angle $\theta_{P'}$ of $\vec{P}_{fb}=\vec{P}_{bf}$ is 90$^{\circ}$, as required by momentum conservation. The effective dielectric constants $\epsilon_{P}$ and $\epsilon_{P'}$ of $\vec{P}_{ff(bb)}$ and $\vec{P}_{fb}$, respectively, are given by  $\epsilon^{1/2}_{P(P')}=\sin{\theta_{\textrm{V}
}}/\sin{\theta_{P(P')}}=\sin{\theta_{\textrm{V}}}/\sin{\theta^{\omega}_{G}}$. 

The measured SH signal is given by the intensity: $I_{\textrm{V}}=\frac{1}{2}\epsilon_0 c\left|E_{\textrm{V}}\right|^2$, where $E_{\textrm{V}}=E_{\textrm{V},b}$ and $\epsilon_0$ is the vacuum permitivity. Matching $E_n$ and $H_n$ across each of the three interfaces yields six equations from which we can determine the six unknown fields $E_{n,f}$ and $E_{n,b}$. The explicit form for $E_{\textrm{V}}$ is
\begin{widetext}
	\begin{equation}
	\label{eq:solution}
	\begin{split}
	E_{\textrm{V}}=&\left\{\left[2\epsilon_{\textrm{G}}^{1/2}\cos{\theta_{\textrm{G}}}M_{ff}+\frac{4\pi\epsilon_{\textrm{G}}^{1/2}\cos{\theta_{\textrm{G}}}(M_{mb}-1)}{\epsilon_{\textrm{P}}-\epsilon_{\textrm{G}}}-\frac{4\pi\epsilon_{\textrm{P}}^{1/2}\cos{\theta_{\textrm{P}}}(M_{mb}+1)}{\epsilon_{\textrm{P}}-\epsilon_{\textrm{G}}}
\right]\cdot P_{ff}+\left[{\vphantom{\frac{\epsilon_{\textrm{G}}^{1/2}}{\epsilon_{\textrm{P}}}}}-2\,\epsilon_G^{1/2}\cos{\theta_{\textrm{G}}}M_{bb}+ \right. \right. \\
	&\left.\left.\frac{4\pi\epsilon_{\textrm{G}}^{1/2}\cos{\theta_{\textrm{G}}}(M_{mb}-1)}{\epsilon_P-\epsilon_{\textrm{G}}}+\frac{4\pi\epsilon_{\textrm{P}}^{1/2}\cos{\theta_{\textrm{P}}}(M_{mb}+1)}{\epsilon_{\textrm{P}}
-\epsilon_{\textrm{G}}}\right]\cdot P_{bb}+\left[-2\epsilon_{\textrm{G}}^{1/2}\cos{\theta_{\textrm{G}}}M_{fb}+\frac{8\pi\epsilon_{\textrm{G}}\cos{\theta_{\textrm{G}}}(M_{mb}-1)}{\epsilon_{\textrm{P}'}-\epsilon_{\textrm{G}}}\right]\cdot P_{fb}\right\}\\
	&\times\left[\epsilon_{\textrm{G}}^{1/2}\cos{\theta_{\textrm{G}}}(M_{mb}-1)+\epsilon_{\textrm{V}}^{1/2}\cos{\theta_{\textrm{V}
}}(M_{mb}+1)\right]^{-1}
	\end{split}
	\end{equation}
\end{widetext}
where 
\begin{equation*}
\begin{split}
M_{ff}=&\frac{(1+T)\epsilon_{\textrm{P}}^{1/2}\cos{\theta_{\textrm{P}}}-(1-T)\epsilon_{\textrm{O}}^{1/2}\cos{\theta_{\textrm{O}}}}{(1+T)\epsilon_{\textrm{G}}^{1/2}\cos{\theta_{\textrm{G}}}-(1-T)\epsilon_{\textrm{O}}^{1/2}\cos{\theta_{\textrm{O}}}}\times \\ 
&\frac{4\pi}{\epsilon_{\textrm{P}}-\epsilon_{\textrm{G}}}\exp\left[-i(\phi_{\textrm{G}}-\phi_{\textrm{P}})\right]\\
\\
M_{bb}=&\frac{(1+T)\epsilon_{\textrm{P}}^{1/2}\cos{\theta_{\textrm{P}}}+(1-T)\epsilon_{\textrm{O}}^{1/2}\cos{\theta_{\textrm{O}}}}{(1+T)\epsilon_{\textrm{G}}^{1/2}\cos{\theta_{\textrm{G}}}-(1-T)\epsilon_{\textrm{O}}^{1/2}\cos{\theta_{\textrm{O}}}}\times \\ 
&\frac{4\pi}{\epsilon_{\textrm{P}}-\epsilon_{\textrm{G}}}\exp\left[-i(\phi_{\textrm{G}}+\phi_{\textrm{P}})\right],\\
\\
M_{mb}=&\frac{(1+T)\epsilon_{\textrm{G}}^{1/2}\cos{\theta_{\textrm{G}}}+(1-T)\epsilon_{\textrm{O}}^{1/2}\cos{\theta_{\textrm{O}}}}{(1+T)\epsilon_{\textrm{G}}^{1/2}\cos{\theta_{\textrm{G}}}-(1-T)\epsilon_{\textrm{O}}^{1/2}\cos{\theta_{\textrm{O}}}}\times\\
&\exp\left(-2i\phi_{\textrm{G}}\right),\\
\\
M_{fb}=&\frac{(1-T)\epsilon_{\textrm{O}}^{1/2}\cos{\theta_{\textrm{O}}}}{(1+T)\epsilon_{\textrm{G}}^{1/2}\cos{\theta_{\textrm{G}}}-(1-T)\epsilon_{\textrm{O}}^{1/2}\cos{\theta_{\textrm{O}}}}\times \\
&\frac{8\pi}{\epsilon_{\textrm{P}'}-\epsilon_{\textrm{G}}} \exp(-i\phi_{\textrm{G}}),\\
\\
T=&\frac{\epsilon_{\textrm{O}}^{1/2}\cos\theta_{\textrm{O}}
-\epsilon_{\textrm{S}}^{1/2}\cos\theta_{\textrm{S}}}{\epsilon_{\textrm{O}}^{1/2}\cos\theta_{\textrm{O}}
+\epsilon_{\textrm{S}}^{1/2}\cos\theta_{\textrm{S}}}\exp(2i\phi_{\textrm{O}}),
\end{split}
\end{equation*}
and the nonlinear polarization terms $P_{ff}$, $P_{bb}$, and $P_{fb}$ are given by Eq. \ref{eqn:P_nls}. For a given $\tensor{\chi}^{(2)}$, these can be calculated easily by solving the corresponding interference problem for the fundamental fields in the multilayer system.

\begin{figure}[hbtp!]
	\includegraphics[width=0.9\columnwidth]{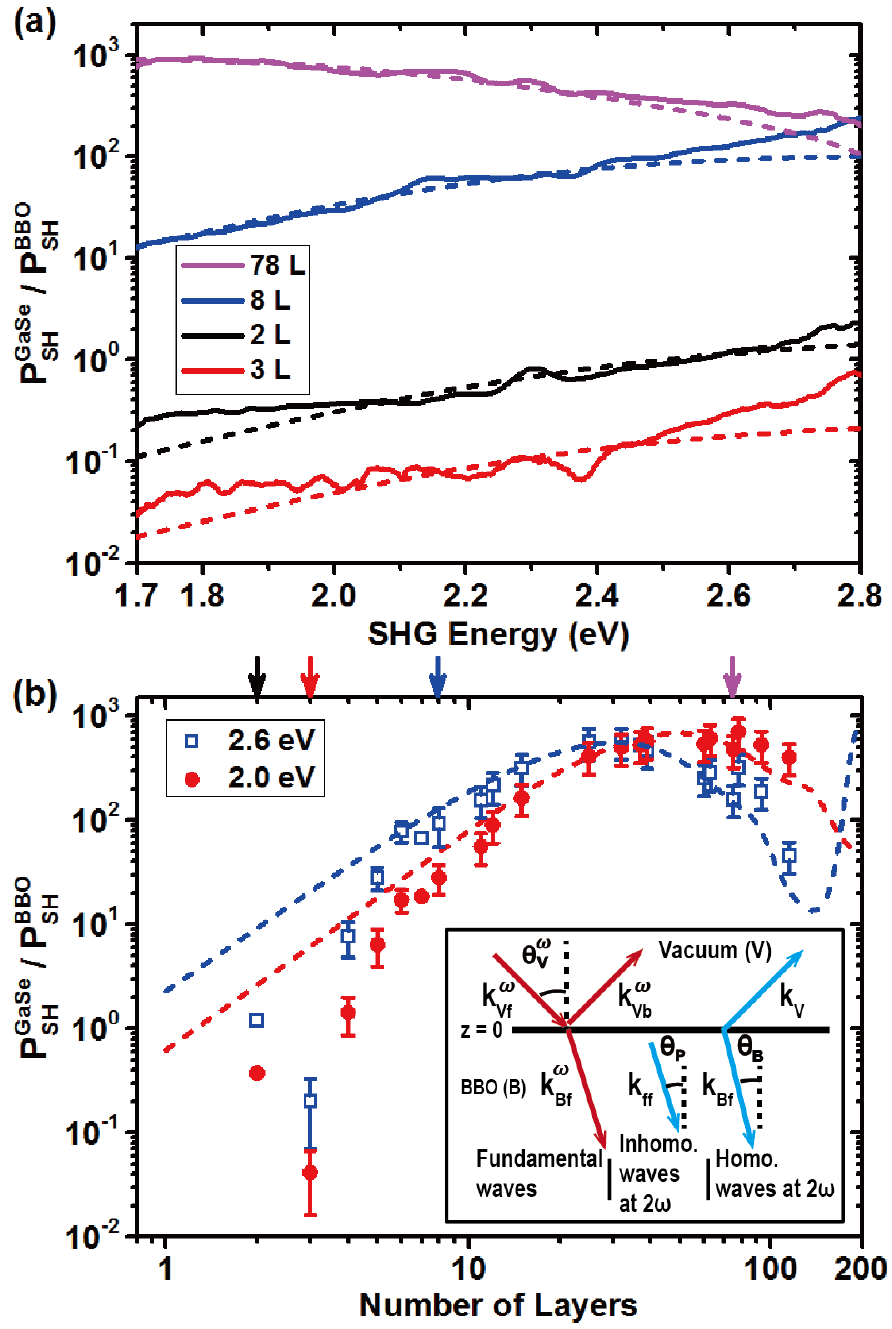}
	\caption{Ratio between the reflected SHG power from GaSe nanoslabs and a thick BBO crystal. (a) Representative curves of the frequency-dependent ratio of the SHG power from 2~L, 3~L, 8~L and 78~L GaSe nanoslabs. The dashed curves are fits to the SHG spectrum from 1.7 eV to 2.4 eV, according to the model in the text. (b) Layer-dependent ratio of the SHG power at the SHG energies $\hbar\omega_{\textrm{SHG}}=2.0$ eV (red dots) and 2.6 eV (open blue squares), with uncertainties determined as in Fig. \ref{fig3_shg_layer}a. The arrows indicate the thickness of samples whose frequency-dependent SHG power are shown in (a). The dashed red and blue curves are fits to the data for thickness $\geq$ 10~L. By comparing to the well known nonlinear optical coefficient of BBO ($d_{22}=2.2$~pm/V), the fit reveals $\big|\chi^{(2)}\big|=80\pm18$ pm/V for $\hbar\omega_{\textrm{SHG}}=$2.0~eV and 2.6~eV. The inset in (b) illustrates the fields involved in calculating the reflected SHG from the BBO crystal.}\label{fig4}
\end{figure}

Given a constant incident fluence and known refractive index\cite{le-toullec1977}, we can determine the value of $\big|\chi^{(2)}\big|$ by fitting the measured layer-dependent SHG power with the above model. For thickness $\geq$ 10~L, we find $\big|\chi^{(2)}\big|=78\pm17$~pm/V. The SHG power decreases by a factor of 42 from 6~L to 3~L, faster than the quadratic layer-dependence expected for a layer-independent value of $\chi^{(2)}$. For nanoslabs with the same number of layers, we find that the sample-to-sample variance of the SHG power is typically $\pm 50\%$. Our results suggest that $\big|\chi^{(2)}\big|$ drops from 6~L to 3~L by a factor of 3 and increases from 3~L to 2~L by a factor of 2.0$\pm$0.7. For 2~L and 1~L, the values of $\big|\chi^{(2)}\big|$ are similar within the sample-to-sample variance. 

A super-quadratic growth in the SHG intensity with layer thickness for few-layer GaSe was previously attributed to a change from $\epsilon$-stacking to $\beta$-stacking\cite{jie2015}. However, $\beta$-GaSe is only centrosymmetric for an even number of layers. In systems, such as hexagonal BN or \ce{MoS2}, displaying centrosymmetry only for even numbers of layers, dramatic differences in the SHG intensity are observed between thin samples with $n$ and $n+1$ layers\cite{li2013}. In light of the absence of such an alternation in SHG intensity with layer thickness in our few-layer samples, we rule out a stacking change from $\epsilon$-stacking to $\beta$-stacking as the primary source of the decrease in SH efficiency in few-layer GaSe. The deviation of the measured values of $\big|\chi^{(2)}\big|$ from the model based on constant $\chi^{(2)}$ occurs for thickness $\lesssim$7~L, where the electronic band structure is calculated to increase dramatically from its bulk value \cite{zhuang2013,do2015,li2015}. This suggests that the reduction of $\big|\chi^{(2)}\big|$ in few-layer GaSe may be a signature of such changes in the band structure.

To further explore the few-layer anomaly in the SHG signal, we measure the frequency-dependent SHG of nanoslabs from 2~L to $>100$~L. To reduce experimental uncertainties in absolute determination of the nonlinear susceptibility, we determine the ratio of the SHG power from GaSe to that from a BBO crystal. The SH fields are determined similarly to the case of GaSe, except that the presence of a single interface greatly simplifies the model, as shown in the inset of Fig.~\ref{fig4}b. The reflected SHG field from a thick BBO crystal, 
$E_{\textrm{V}}$ is given by
\begin{equation}
\label{eq:SHG_BBO}
E_{\textrm{V}}=
\frac{-4\pi P^{(2)}_{\textrm{B}}}{(\epsilon_{B}^{1/2}\cos{\theta_{B}}+\cos{\theta_{\textrm{V}}})(\epsilon_{B}^{1/2}\cos{\theta_{B}}+\epsilon_{\textrm{P}}^{1/2}\cos{\theta_{P}})}
\end{equation}
where 
$\vec{P}_B=\vec{P}_{\textrm{B},ff}=\tensor{\chi}^{(2)}:\vec{E}_{\textrm{B},f}^{\omega}\vec{E}_{\textrm{B},f}^{\omega}$, 
$\theta_{\textrm{V}}$ is the incident angle of the fundamental field, $\theta_B$ and $\epsilon_B$ are respectively the refracted angle and dielectric constant of the SH field in the BBO crystal and are related to $\theta_{\textrm{V}}$  by $\sin\theta_{\textrm{V}}=\epsilon_B^{1/2}\sin{\theta_B}$, and $\theta_P$ is the refracted angle of the inhomogeneous field and is the same as the angle of the fundamental field in the BBO crystal. We show four representative spectra of the frequency-dependent ratio of the SHG power in Fig.~\ref{fig4}a. We fit the spectra of the SHG power from 1.7 to 2.4 eV with a spectrally constant value of $\chi^{(2)}$ according to the model described above. The fittings reveal $\big|\chi^{(2)}\big|=$102$\pm17$, 66$\pm11$, 7.2$\pm1.8$ and 28$\pm6.4$ pm/V for samples of 78, 8, 3 and 2 layers, respectively. The mismatch of 78- and 8-L samples at high SH energy might be due to enhancement of $\chi^{(2)}$ by high-energy resonances. 

We do not observe any features indicative of excitonic resonances at the bandgap in the SHG spectra of Fig.~\ref{fig4}a. This is in contrast to the case of \ce{WSe2} monolayers, in which SHG has been observed to increase by three orders of magnitude near the two-photon excitonic resonance \cite{wang2015}. The absence of excitonic enhancement in GaSe may be attributed to the weak absorption for light with electric field $\vec{E}$ perpendicular to the crystalline $c$-axis \cite{levine1974,le-toullec1977}. We note that previous studies of bulk GaSe only revealed a weak two-photon excitonic resonance in the SHG power after accounting for the frequency-dependent optical absorption \cite{hirlimann1989}. Even a low-temperature study~\cite{allakhverdiev1995} only showed an increase of the SHG power by a factor of 2 at the one-photon excitonic resonance. This suggests that for wave propagation along the $c$-axis, the SHG response in GaSe is dominated by transitions other than those giving rise to band-edge absorption. This does not rule out the possibility that the reduction in SHG from few-layer GaSe is associated with predicted changes in the highest valence band and lowest conduction band \cite{zhuang2013,do2015,li2015}. For example, the primary contributions to the SHG response could come from transitions between one of these bands and higher conduction or lower valence bands. However, further calculations would be needed to understand the source of the SHG response.

In Fig.~\ref{fig4}b, we show experimental data and fitting of the layer-dependent ratio of the SHG power from GaSe to the SHG power from BBO at the SHG energies $\hbar\omega_{\textrm{SHG}}=2.0$~eV and 2.6~eV. Our model reveals that for both $\hbar\omega_{\textrm{SHG}}=2.0$~eV and 2.6~eV, $\big|\chi^{(2)}\big|=80\pm18$ pm/V, consistent with the result derived by the absolute measurement at $\hbar\omega_{\textrm{SHG}}=3.16$~eV. The similar 
value of $\big|\chi^{(2)}\big|$ at $\hbar\omega_{\textrm{SHG}}$ from 2.0 eV to 3.16 eV again indicates that any excitonic contribution to the nonlinear optical susceptibility at room temperature is weak. 

\section{Conclusions}
We have measured the room-temperature second-harmonic response of GaSe from monolayer to $>$100 layers for sub-bandgap fundamental photon energies but second-harmonic photon energies from 1.7 to 3.1 eV, i.e., tuning the second-harmonic across the bandgap. For crystals of thickness $>$10 layers, we obtain a value of the second-order susceptibility of $\big|\chi^{(2)}\big|=80\pm18$~pm/V for a fundamental photon energy below 1.3 eV, which is close to that reported for the bulk. Deviations from the bulk second-order susceptibility $\chi^{(2)}$ are observed only for thicknesses $\lesssim$~7 layers, at which thicknesses $\chi^{(2)}$ is suppressed. No signatures of two-photon excitonic resonances are observed, which is consistent with the weak excitonic absorption for electric fields polarized perpendicular to the $c$-axis.

\begin{acknowledgments}
We thank Brage Golding for discussions. This work was supported by NSF grant DMR-09055944 as well as start-up funding and the Cowen endowment at Michigan State University. This research has used the W. M. Keck Microfabrication Facility.
\end{acknowledgments}

%\bibliography{JAM_SHG_lib}
%\bibliography{/Users/cwlai/Dropbox/Bib/lai_lib}% Produces the bibliography via BibTeX.
%

\end{document}